\documentclass[aps,twocolumn,superscriptaddress]{revtex4-1}
\usepackage{amsmath}    
\usepackage{bm}
\usepackage{soul}
\usepackage{graphicx}   
\usepackage{verbatim}   
\usepackage{color}      
\usepackage{hyperref}   
\begin{document}

\title{Tuning magnetic antiskyrmion stability in tetragonal inverse Heusler alloys}

\author{Daniil A. Kitchaev}
\email{dkitch@alum.mit.edu}
\affiliation{Materials Department, University of California, Santa Barbara, California 93106, USA}

\author{Anton Van der Ven}
\affiliation{Materials Department, University of California, Santa Barbara, California 93106, USA}

\begin{abstract}
The identification of materials supporting complex, tunable magnetic order at ambient temperatures is foundational to the development of new magnetic device architectures. 
We report the design of Mn${}_2 XY$ tetragonal inverse Heusler alloys that are capable of hosting magnetic antiskyrmions whose stability is sensitive to elastic strain.
We first construct a universal magnetic Hamiltonian capturing the short- and long- range magnetic order which can be expected in these materials.
This model reveals critical combinations of magnetic interactions that are necessary to approach a magnetic phase boundary, where the magnetic structure is highly susceptible to small perturbations such as elastic strain.
We then computationally search for quaternary Mn${}_2 (X_1, X_2) Y$ alloys where these critical interactions may be realized and which are likely to be synthesizable in the inverse Heusler structure.
We identify the Mn${}_2$Pt${}_{1-z} X_z$Ga family of materials with $X = $ Au, Ir, Ni as an ideal system for accessing all possible magnetic phases, with several critical compositions where magnetic phase transitions may be actuated mechanically.
\end{abstract}

\maketitle

A substantial component of spintronic device development is the discovery of materials that are capable of hosting exotic spin textures over precisely tuned field and temperature ranges\cite{Bader2010}.
While spin textures can be controlled by magnetic fields, dynamically coupling these magnetic phases to other variables such as electric fields or mechanical perturbations allows for new control paradigms and device architectures\cite{Yang2021}.
Magnetic skyrmion and antiskyrmion textures have in particular attracted attention due to their combination of thermodynamic stability, unique topological properties, and efficient transport behavior\cite{Bogdanov1994, Roessler2006, Muhlbauer2009, Jonietz12010, Yu2012, Fert2017}. 
A number of bulk material systems capable of hosting equilibrium skyrmions\cite{Muhlbauer2009, Tokunaga2015, Kezsmarki2015, Janson2016, Bordacs2017, Fujima2017, Schueller2020} or antiskyrmions\cite{Nayak2017, Jena2019} have been discovered, and recent reports have indicated that certain materials may support both topologies as metastable states\cite{Peng2020, Jena2020}.
However, tuning the geometry and stability windows of these topological phases, at the synthesis stage or \emph{in situ}, remains a challenge. 
This is due to the lack of a theoretical understanding of flexible material systems that are capable of hosting (anti)skyrmion phases at room temperature, as well as the irreversibility of the structural deformations typically necessary to elicit a substantial magnetic response\cite{Levin2021, Schueller2020}.

An attractive model system for realizing chemically- and mechanically- tunable (anti)skyrmion states are the tetragonal inverse Heusler alloys\cite{Felser2013, Nayak2017, Jena2019}.
These materials have the Mn${}_2 X Y$ chemical formula where the $X$ sublattice generally consists of late transition metal elements and $Y =$ Ga, Sn, In\cite{Felser2013, Wollmann2015}.
Below their martensitic transformation temperature, they possess $D_{2d}$ symmetry.
This symmetry is compatible with the formation of thermodynamically-stable antiskyrmions\cite{Hoffmann2017, Leonov2017}, while metastable skyrmions can be nucleated with an appropriate history of applied magnetic fields\cite{Peng2020, Jena2020}.
Critically, this symmetry also ensures that the topological phases may remain stable from 0 K to $T_c$ (Curie temperature)\cite{Kitchaev2020, Bogdanov1994, Leonov2017}, which in these materials is often well above room temperature\cite{Wollmann2015}.
Furthermore, the Heusler alloys allow for immense chemical flexibility, which has been previously used to tune their structural\cite{Liu2005}, electronic\cite{Chadov2015, Huh2015} and magnetic\cite{Winterlik2012,decolvenaere2019modeling} properties.
The combination of chemical flexibility in the $X$ and $Y$ sublattices and thermal stability of the topological magnetic states means that the topological magnetism seen in Mn${}_2 X Y$ inverse Heusler alloys may in principle be tuned and observed at room temperature, as is necessary for device applications.

Here, we implement a general design strategy for realizing chemically and mechanically tunable antiskyrmions using the Mn${}_2 X Y$ tetragonal inverse Heuslers as a model system.
We first derive a universal model for the short- and long- range magnetic order of materials with the inverse Heusler structure in terms of computable magnetic interactions.
We then enumerate known and hypothetical Mn${}_2 X Y$ inverse Heusler materials and characterize the impact of varying chemistry on the magnetic interactions and chemical stability of the alloys.
We show that with an appropriate choice of composition on the $X$ and $Y$ sublattices, one can realize all possible magnetic phases and create materials where magnetic phase transitions may be actuated with small, purely elastic mechanical perturbations.
Finally, we identify Mn${}_2$Pt${}_{1-z}X_z$Ga with $X=$Au, Ir, Ni and $z \approx 0.1-0.2$ as an ideal system for realizing this behavior, combining flexible room-temperature magnetism with chemical and structural stability. 

\section*{Results}

\begin{figure}[t!]
\includegraphics[width=0.48\textwidth]{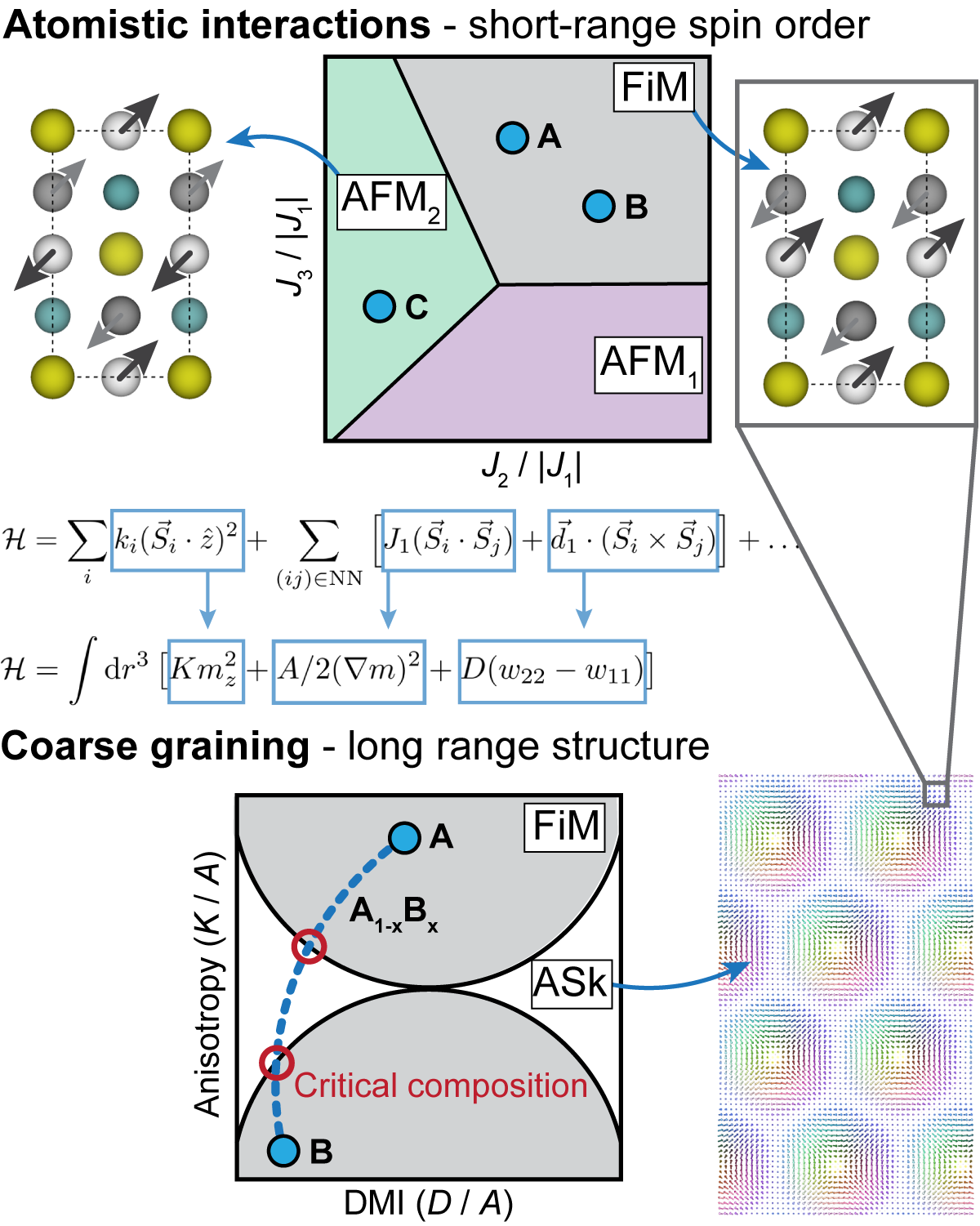}
\caption{\label{fig:methods} Schematic summary of our materials design strategy for obtaining tunable antiskyrmion behavior. We first map all hypothetical compounds $A$, $B$, $C$ to interaction parameters $J_1, J_2, \hdots$ of a quasi-classical atomistic Hamiltonian to determine their short-range spin order. We repeat this analysis at longer length scales by coarse-graining the magnetocrystalline anisotropy $K$ and Dzyaloshinskii-Moriya interaction $D$ (DMI). We identify the parameter space where antiskyrmions (ASk) may be expected and construct alloys $A_{1-x}B_x$ which fall in the region of ASk stability. Finally, we identify critical compositions $x_c$ falling on magnetic phase boundaries as compounds where the magnetic phase transition may be actuated by small perturbations, e.g. reversible elastic strain.
}
\end{figure}

Our approach to realizing chemically- and mechanically- tunable antiskyrmion states is to relate the form of the magnetic phase diagram to variations in atomistic magnetic interactions, and then characterize how these interactions may be tuned by chemical changes and elastic perturbations.
This approach is shown schematically in Figure \ref{fig:methods}.
We first construct an atomistic quasi-classical spin Hamiltonian applicable to all tetragonal inverse Heuslers, accounting for local exchange, Dzyaloshinskii-Moriya (DMI) and anisotropy interactions.
Using this model, we parametrically enumerate the local spin structures that can be stabilized by various combinations of exchange strengths.
Next, we coarse-grain these atomistic interactions to produce a continuum free energy functional that enables a parametric  exploration of long-range magnetic structures such as antiskyrmions.
We deduce the magnetic behavior of candidate materials $A$, $B$, $C$ by fitting their interaction parameters to density functional theory (DFT) data.
We then construct alloys $A_{1-x}B_x$ between compatible materials $A$ and $B$ that share the same local spin order.
In the $A_{1-x}B_x$ alloy, coarse-grained magnetic interactions vary continuously with composition, making it possible to identify critical compositions $x_c$ that reside on magnetic phase boundaries where magnetic phase transitions may be actuated by small perturbations such as reversible elastic strain.

\subsection*{Short- and long-range magnetic order in tetragonal inverse Heuslers}

\begin{figure}[t!]
\includegraphics[width=0.48\textwidth]{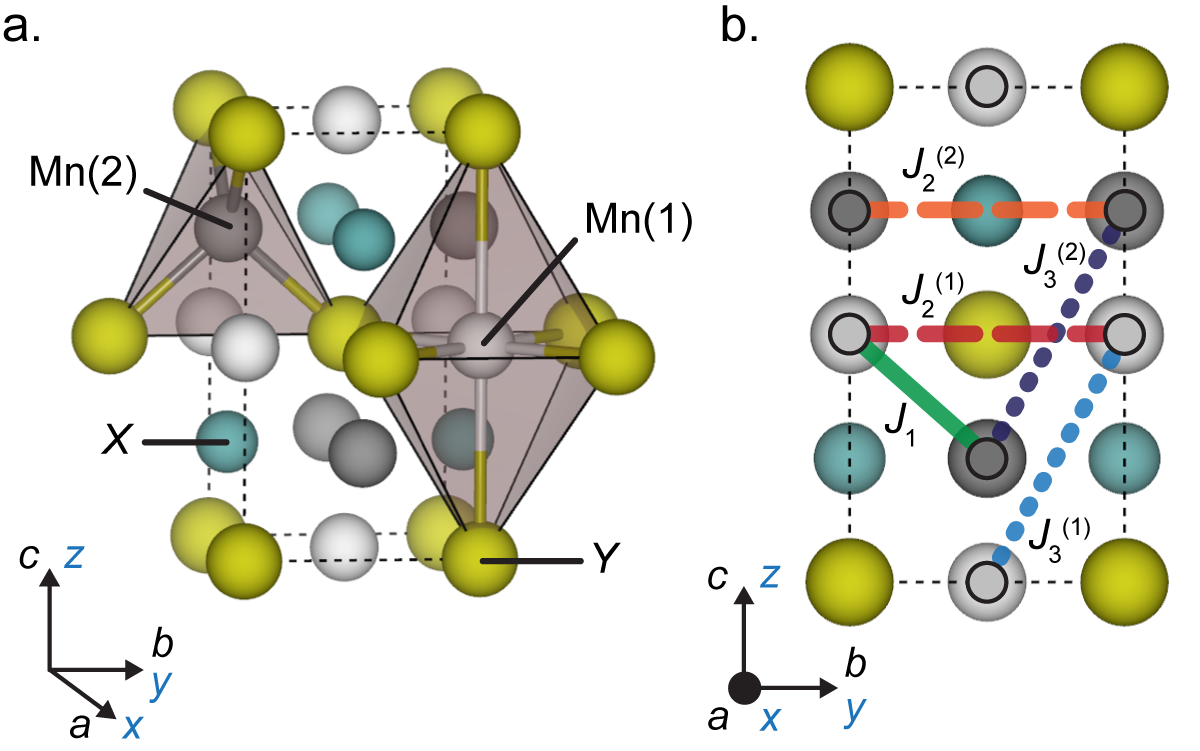}
\caption{\label{fig:structure} Idealized structure of a Mn${}_2 XY$ tetragonal inverse Heusler alloy.
\textbf{a.} Distinct crystallographic sites which define four interpenetrating face-centered cubic lattices.
\textbf{b.} Minimal magnetic interaction model for a prototypical Mn${}_2 XY$ tetragonal inverse Heusler. 
$J_1$ represents the coupling of the Mn(1) and Mn(2) sublattices. 
$J_2^{(1)}$ and $J_3^{(1)}$ represent the in-plane and out-of-plane interactions respectively within the Mn(1) sublattice, and $J_2^{(2)}$ and $J_3^{(2)}$ for the Mn(2) sublattice.
}
\end{figure}

The Mn${}_2 X Y$ tetragonal inverse Heusler alloys are defined by the idealized crystal structure shown in Figure \ref{fig:structure}a.
This structure consists of four tetragonally distorted interpenetrating face-centered-cubic sublattices.
Two of these sublattices, Mn(1) and Mn(2), have localized magnetic moments in the range of 2-3 $\mu_{\text{B}}$ per atom.
The $Y$ sublattice generally contains one of Ga, Sn, or In and is non-magnetic.
The $X$ sublattice can be occupied by a range of late transition-metal elements, with previously reported compounds having $X = $ Fe, Co, Ni, Rh, Pd, or Pt\cite{Wollmann2015, Kreiner2014, Faleev2017, Winterlik2012}.
In this study, we supplement these elements with other transition metals which could potentially be doped onto the $X$ sublattice: Ru, W, Os, Ir, Au\cite{Wollmann2015}
However, we exclude Fe as it introduces a large magnetic moment on the $X$ sublattice and cannot be treated with the same magnetic model as systems with non-magnetic $X$ elements.
As both the chemical stability and the degree of chemical order vary substantially between these chemistries, we will discuss which compositions are most likely to be synthetically accessible in a later section.

\begin{figure*}[t!]
\includegraphics[width=\textwidth]{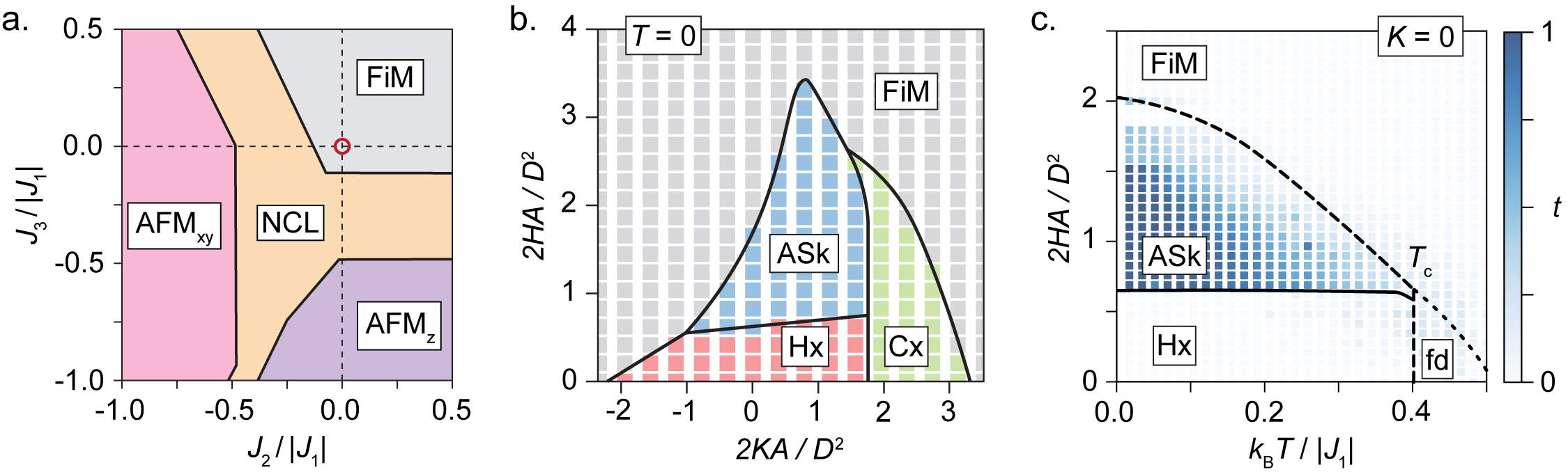}
\caption{\label{fig:mag_pd} Equilibrium phases given by the minimal magnetic Hamiltonian for the Mn${}_2 XY$ tetragonal inverse Heusler structure.
\textbf{a.} Local spin configurations governed by the relative strength of $J_1$-$J_2$-$J_3$ exchange interactions, including collinear ferrimagnetic (FiM) and antiferromagnetic (AFM${}_{xy}$, AFM${}_{z}$) phases, as well as a region of frustrated non-collinear order (NCL).
\textbf{b.} Long-range phases generated as modulations of the FiM order at low-$T$, which include spin helices (Hx), antiskyrmion lattices (ASk) and conical helices (Cx).
Long-range structure is governed by the relative strength of uniaxial anisotropy $K$, Dzyaloshinskii-Moriya coupling $D$, spin-stiffness $A$ and applied field $H$ along the $c$-axis.
The phase diagram is evaluated for $J_2 = J_3 = 0$ (red circle in \textbf{a.}) and an equilibrium helical wavelength of 24 unit cells.
\textbf{c.} Extension of the $K = 0$ region of the long-range phase diagram to finite temperature.
Color denotes the expected number of antiskyrmions per 24x24 unit cell as measured by the topological index density $t$. 
Solid lines denote first-order phase transitions while dotted lines denote second-order or continuous phase boundaries.
$T_c$ denotes the Curie temperature and fd refers to the fluctuation-disordered Brazovskii region\cite{Janoschek2013, Kitchaev2020}.
}
\end{figure*}

We represent the magnetic behavior of Mn${}_2 X Y$ inverse Heuslers with a combination of exchange interactions, which are the dominant energy scale and control the local spin structure, and coarse-grained spin-orbit effects that control the long-range modulation of the local spin structure.
We consider the atomic exchange interactions in conventional Heisenberg model form,  $\mathcal{H}_{\text{exchange}} = \sum_{ij \in \alpha} J_{\alpha} (-\vec{S}_i \cdot \vec{S}_j)$
 where the summation includes couplings up to the 3rd nearest neighbor as shown in Figure \ref{fig:structure}b.
$J_1$ represents the strongly antiferromagnetic direct exchange between the Mn(1) and Mn(2) sublattices, while $J_2$ and $J_3$ capture the weaker interactions within the two sublattices.
To further simplify the model, we set $J_2 = J_2^{(1)} = J_2^{(2)}$ and $J_3 = J_3^{(1)} = J_3^{(2)}$ so that the geometrically-identical interactions within the Mn(1) and Mn(2) sublattices are assumed to have the same interaction strength. 
The complete form of this spin Hamiltonian is given in Supplementary Data 1.
Despite the simplicity of this model, we find that it is sufficient to capture the energetics of collinear spin configurations in the Mn${}_2 X Y$ compounds considered in this work, reproducing both the ground state and excited state spectrum as computed with density functional theory (DFT).
The results of this fitting procedure and the correspondence between the model and the electronic structure data is quantified in Supplementary Data 2 and 3.

The competition between the exchange interactions $J_1$, $J_2$ and $J_3$ gives rise to several local spin orderings, as shown in Figure \ref{fig:mag_pd}a.
When $J_2$ and $J_3$ are ferromagnetic, or negligible compared to $J_1$, the spins adopt a ferrimagnetic structure (FiM) with the Mn(1) and Mn(2) sublattices antialigned with each other.
This structure has a net moment as the local moment on Mn(1) is typically larger than that on Mn(2).
Antiferromagnetic $J_2$ and $J_3$ interactions frustrate this order and can lead to a region of non-collinear order (NCL), or collinear antiferromagnetic structures with spins either alternating in the $xy$ plane or along the $z$ axis (AFM${}_{xy}$ and AFM${}_z$ respectively).
Of these structures, we focus on the ferrimagnetic phase as it is the only spin structure with a net magnetic moment at low temperature and field.

The long-range magnetic texture is defined by a gradual rotation of the local spin structure driven by the Dzyaloshinskii-Moriya component of spin-orbit coupling and suppressed by the magnetocrystalline anisotropy and spin-stiffness.
These phenomena are conventionally described by a coarse-grained magnetic Hamiltonian for the $D_{2d}$ point group\cite{Kitchaev2018,Abert2019}:
\[
\mathcal{H} = \int \mathrm{d}r^3 \left[ A/2 (\nabla m)^2 + D (w_{22} - w_{11}) + K m_z^2 \right]
\]
where $m$ is the unit vector direction of the local magnetization.
$A$ is the spin-stiffness parameter and represents the coarse-grained exchange strength.
The relationship between $A$ and the atomistic $J_1$, $J_2$, $J_3$ parameters is given in Supplementary Data 4.
$D$ and $w_{kn} = \epsilon_{ijk} m_i \partial m_j / \partial r_n$ represent the strength and form of the coarse-grained Dzyaloshinskii-Moriya interaction, where $\epsilon_{ijk}$ is the Levi-Civita tensor and repeated indices imply summation\cite{Kitchaev2018}.
$K$ parametrizes the uniaxial anisotropy with respect to the crystal axes given in Figure \ref{fig:structure}a.
While higher-order anisotropies are necessary to accurately capture the DFT energetics of some Heusler compounds including the Pt and Ir-based systems considered here, we have found that these terms are never large enough to alter the final magnetic phase diagrams in our analysis.
Here, all spatial dimensions are taken to be in units of the lattice parameter of the conventional structure shown in Figure \ref{fig:structure}a ($a$ for the $xy$ directions,  $c$ for the $z$ direction).

\begin{figure*}[t!]
\includegraphics[width=\textwidth]{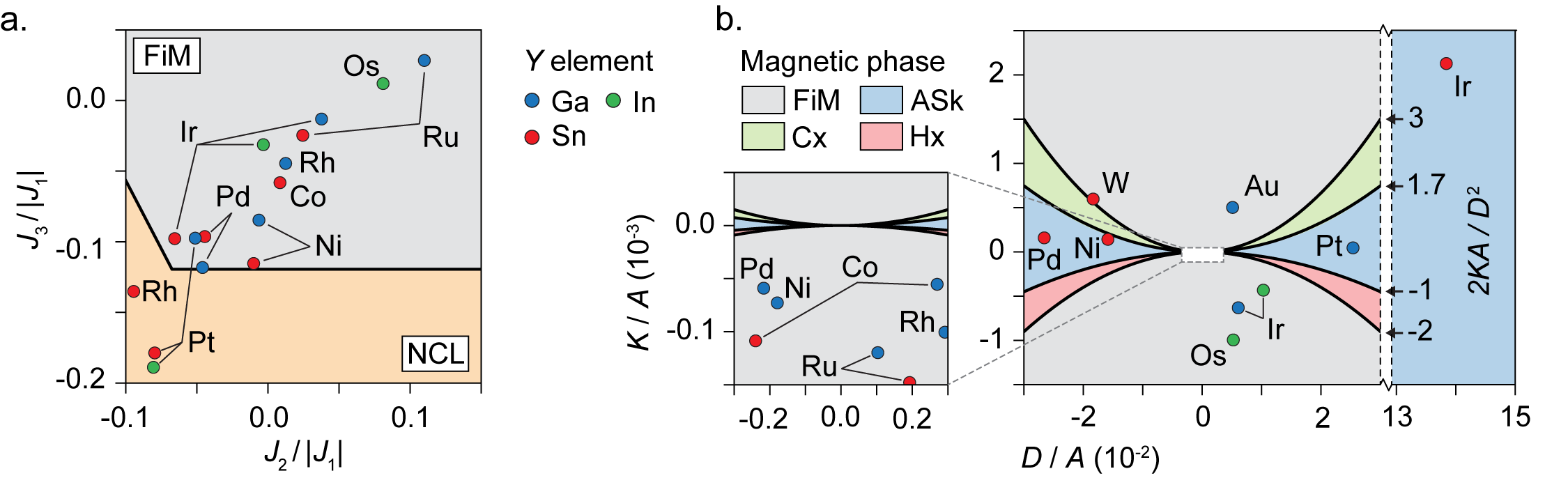}
\caption{\label{fig:mag_chems} Magnetic interactions in ternary Mn${}_2 XY$ inverse Heusler compounds.
\textbf{a.} Local spin order based on competition between exchange interactions, including only those chemistries which favor the tetragonal inverse Heusler structure at the Mn${}_2 XY$ stoichiometry.
\textbf{b.} Long-range spin textures in chemistries favoring local FiM order, based on the relative strengths of the Dzyaloshinskii-Moriya ($D$) and mangetocrystalline anisotropy ($K$) components of spin-orbit coupling.
See caption to Figure \ref{fig:mag_pd} for phase definitions.
Note that $D/A = 2\pi / \lambda$ where $\lambda$ is the equilibrium wavelength of the helical and antiskyrmion phases in units of the basal ($a$) lattice parameter.
}
\end{figure*}

Whether or not the local spin structure develops a long-range texture at equilibrium is determined by the competition between the Dzyaloshinskii-Moriya and magnetocrystalline anisotropy components of spin-orbit coupling ($D/A$ and $K/A$ respectively)\cite{Kitchaev2020, Bogdanov1994}.
These spin textures include spin helices (Hx), spin cones (Cx) and antiskyrmions (ASk) which all have a characteristic energy scale of $D^2/2A$ and form the phase diagram shown in Figure \ref{fig:mag_pd}b in the low temperature limit.
This phase diagram shows that as a function of the normalized anisotropy ($2KA / D^2$) and magnetic field along the $c$-axis ($2HA / D^2$), spin helices and conical structures are stabilized for $-2 \leq 2KA / D^2 \leq 3$.
Antiskyrmions are favored under a small applied field for $-1 \leq 2KA / D^2 \leq 1.7$.
The change in this phase diagram at elevated temperature is shown in Figure \ref{fig:mag_pd}c for the case of vanishing anisotropy $K$ and $J_2 = J_3 = 0$.
The helical and antiskyrmion phases persist at all temperatures up to $T_c$ with minimal change in the phase boundary between them, although the maximum field at which antiskyrmions are stable decreases.
Variation of $J_2$ and $J_3$ within the FiM region do not alter the overall shape of this phase diagram, but do significantly rescale $T_c$ as shown in Supplementary Data 4.

\subsection*{Magnetic structure and chemical stability of Mn${}_2 X Y$ tetragonal inverse Heuslers}

We now examine where known and hypothetical ternary Mn${}_2 X Y$ inverse Heuslers fall on the magnetic phase diagrams shown in Figure \ref{fig:mag_pd}.
In Figure \ref{fig:mag_chems}a we plot the exchange interactions in a range of compounds and deduce their local spin structure, focusing only on those compositions that thermodynamically favor the tetragonal inverse Heusler structure at the Mn${}_2 X Y$ composition.
The majority of these compounds fall in the FiM region, with frustrated non-collinear order expected in Mn${}_2$PtSn, Mn${}_2$PtIn and Mn${}_2$RhSn consistent with experimental reports\cite{Liu2018, Nayak2012, Felser2013, Meshcheriakova2014, Giri2020}.
For the remaining materials that favor FiM order, we compare the coarse-grained spin-orbit coupling to the regions where helical, conical, or antiskyrmion long-range phases can be expected.
Here we also include the hypothetical compounds Mn${}_2$AuGa and Mn${}_2$WSn for which the inverse Heusler chemical order is metastable.
As can be seen in Figure \ref{fig:mag_chems}b, most compositions are easy-axis ferrimagnets (FiM with $K < 0$), with only the hypothetical Mn${}_2$AuGa and Mn${}_2$WSn materials falling in the easy-plane ferrimagnet region (FiM, $K > 0$).
Non-collinear spin textures can be expected in Mn${}_2$PtGa, Mn${}_2$IrSn, Mn${}_2$PdSn and Mn${}_2$NiSn, where the Dzyaloshinskii-Moriya interaction is sufficiently large to fall in the antiskyrmion stability region ($-1 \leq 2KA / D^2 \leq 1.7$).

Of the various compounds mapped out in Figure \ref{fig:mag_chems}, we focus on Mn${}_2X$Ga for $X$ = Pt, Ni, Ir, as they are the most likely to be synthesizable at equilibrium as stoichiometric, well-ordered inverse-Heusler compounds.
The synthesis of any inverse Heusler compound can be challenging, as the finite-temperature phase diagrams of the binary endpoints are often very complicated and the phase diagrams of the full ternary systems are not known.
For example, the prototypical compound for the tetragonal inverse Heusler structure, Mn${}_3$Ga, forms by a low-temperature peritectic reaction with numerous competing phases that need to be avoided to produce a high-quality material\cite{Hao2020, Balke2007}. 
Furthermore, the formation of chemical order is complicated by the coupling between chemical order and the structural transformation between the high-temperature austenite and low-$T$ martensite phase\cite{Uijttewaal2009, Brown2010, Wollmann2015}.
While assessing the full finite-temperature phase diagrams and ordering kinetics of the chemistries described here is prohibitive, we can evaluate the likelihood that any given Mn${}_2XY$ may be formed by the conventional process of high-temperature mixing followed by a long low-temperature anneal.
We assume that the high-temperature precursor is a disordered state prepared at the correct stoichiometry\cite{Felser2013}. 
As this precursor is cooled, the formation of an ordered product is characterized by a driving force $\Delta E_{\text{order-disorder}}$, which we approximate using the difference in energy between the ordered inverse Heusler product and most favorable disordered state among the common disorder models proposed for these systems (L$2_{1b}$ (Mn(1)/$X$), BiF${}_3$ (Mn(1)/Mn(2)/$X$))\cite{Kreiner2014}.
This ordering reaction competes with phase separation, whose likelihood is correlated with the energy of formation $\Delta E_{\text{formation}}$ of the target compound from competing phases in each Mn-$X$-$Y$ ternary space\cite{Sanvito2017}.
The equilibrium phases used to determine chemical stability are given in Supplementary Data 5.

Figure \ref{fig:stability} charts the driving forces for the competing order-disorder and decomposition reactions for the Mn${}_2XY$ chemistries discussed here, excluding the W-based compounds and Mn${}_2$AuSn as they are exceptionally unstable.
All In-based and most Sn-based compounds have a strong driving force for phase separation and thus are not likely to retain the desired stoichiometry after a long anneal. 
Furthermore, a number of compounds have a minimal driving force to order, or in the case of Pd-based systems and Mn${}_2$AuGa do not favor the ordered states we have considered at all.
The systems which favor the ordered inverse Heusler structure at low-$T$ equilibrium are Mn${}_2 X$Ga for $X = $Ir, Pt, Rh, Ru, Ni, Co and Mn${}_2X$Sn for $X = $Ru, Rh.
From these, we exclude the Ru-based systems and Mn${}_2$RhGa as experimental reports of these compounds indicate that the ordered configuration is difficult to obtain in practice\cite{Kreiner2014}, Mn${}_2$RhSn as it does not favor the locally-collinear FiM phase, and Mn${}_2$CoGa as it does not produce a tetragonal distortion.
We now focus on the remaining synthesizable chemistries to identify combinations which, when alloyed, may generate magnetic phase transitions.

\begin{figure}[t!]
\includegraphics[width=0.4\textwidth]{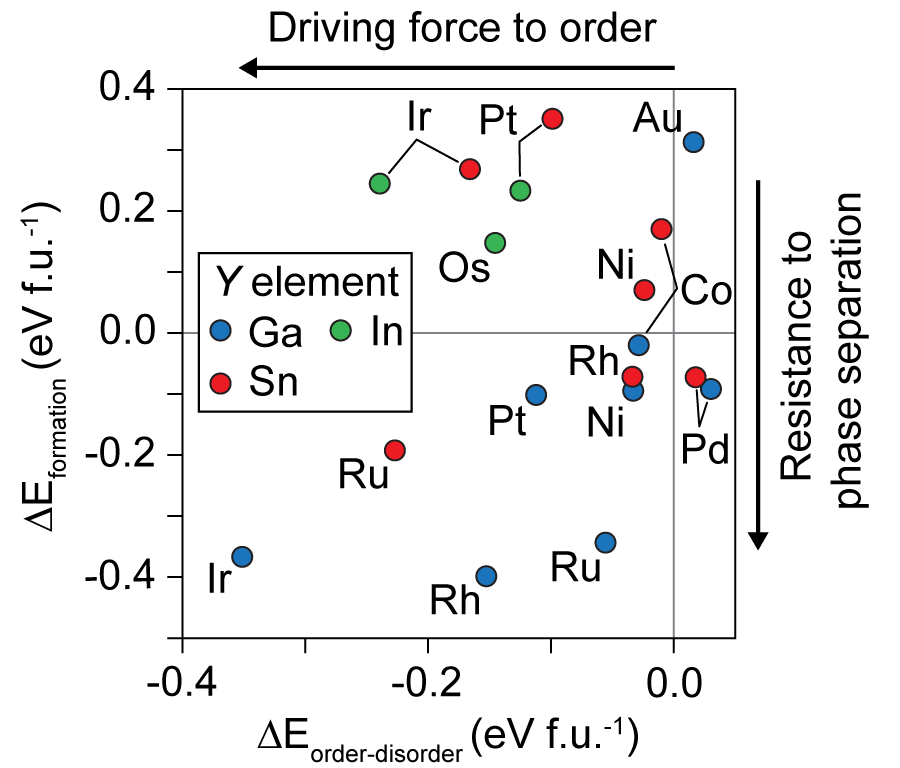}
\caption{\label{fig:stability} Likelihood that an ordered inverse Heusler compound may be obtained by an equilibrium synthesis method at the Mn${}_2 XY$ stoichiometry.
$\Delta E_{\text{order-disorder}}$ measures the driving force for ordering and is defined as the minimum difference in energy between the ordered structure and the common types of disorder observed in inverse Heuslers (L$2_{1b}$ (Mn(1)/$X$), BiF${}_3$ (Mn(1)/Mn(2)/$X$))\cite{Kreiner2014}. 
$\Delta E_{\text{formation}}$ measures the likelihood of phase separation into other phases in the Mn-$X$-$Y$ chemical space and is defined as the difference in zero-$T$ energy between the ordered Mn${}_2 XY$ phase and an equilibrium combination of competing phases given in Supplementary Data 5.}
\end{figure}

\subsection*{Designing tunable magnetism in Mn${}_2$Pt${}_{1-z}X_z$Ga alloys}

\begin{figure*}[t!]
\includegraphics[width=\textwidth]{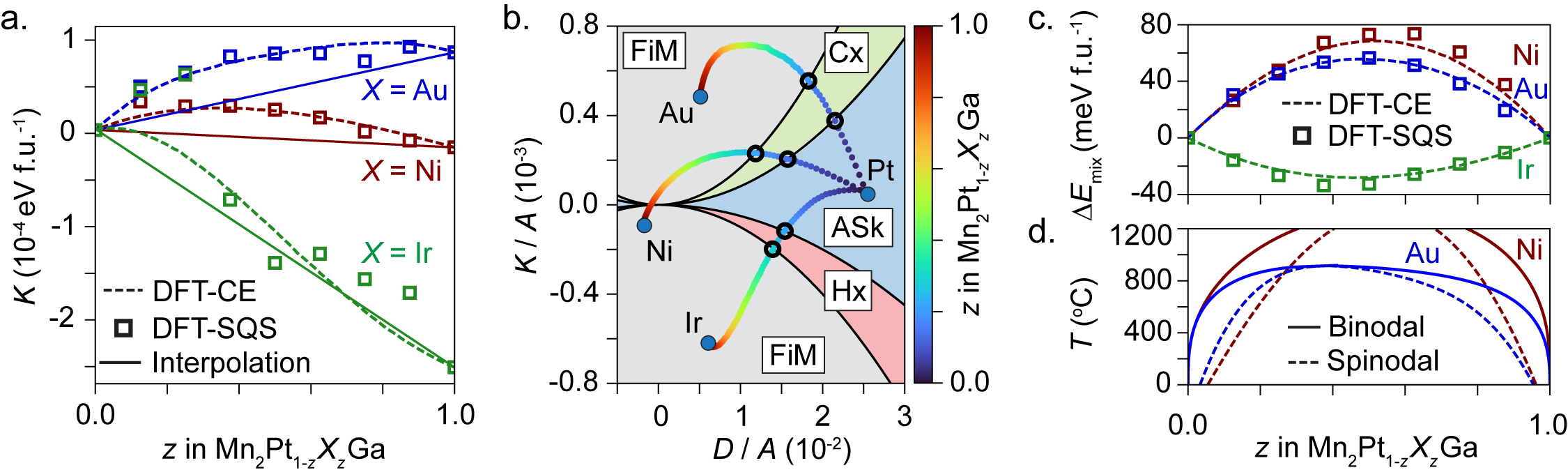}
\caption{\label{fig:pt_doping} Evolution of magnetic interactions and phases in  Mn${}_2$Pt${}_{1-z}$X${}_z$Ga alloys for $X =$ Au, Ir, Ni. \textbf{a.} Evolution of mesoscopic uniaxial anisotropy $K$ with composition, evaluated by linear interpolation between the alloy endpoints, a parametrized cluster expansion (DFT-CE), or explicit calculations of a SQS structure (DFT-SQS). \textbf{b.} Location of various alloy compositions in magnetic phase space, based on the cluster expansion model of anisotropy $K$, and linear interpolation of spin stiffness $A$ and the Dzyaloshinskii-Moriya interaction $D$. Note that $D/A = 2\pi / \lambda$ where $\lambda$ is the equilibrium wavelength of the helical and antiskyrmion phases in units of the basal ($a$) lattice parameter. \textbf{c.} Mixing energy of Mn${}_2$PtGa-Mn${}_2X$Ga evaluated using a parametrized cluster expansion and the SQS methods. \textbf{d.} Binodal and spinodal regions of the miscibility gap in the $X=$ Au, Ni alloys based on the DFT-CE mixing enthalpy given in c., and an ideal entropy model for the $X$ sublattice.}
\end{figure*}

Having enumerated the magnetic behavior and chemical stability of the ternary Mn${}_2XY$ inverse Heuslers, we turn to quaternary alloys in this space to tune magnetic properties and fully explore the magnetic phase diagram shown in Figure \ref{fig:mag_pd}b.
In a solid solution between two compounds Mn${}_2X^{(1)}Y^{(1)}$ and Mn${}_2X^{(2)}Y^{(2)}$ with the same local chemical and spin structure, the coarse-grained magnetic parameters $D$, $K$ and $A$ must vary continuously with composition.
Graphically, this continuous variation means that the magnetic parameters of the alloy will fall on a smooth curve connecting the endpoint compounds in Figure \ref{fig:mag_chems}b.
By alloying two materials which are separated by a magnetic phase boundary in Figure \ref{fig:mag_chems}b, we can switch the magnetic behavior of the alloy between the two phases by varying composition.
Furthermore, we can identify the alloy composition that lies on the magnetic phase boundary to create a material where the magnetic phase transition can be actuated by a small elastic strain.

On the basis of the magnetic interaction parameters shown in Figure \ref{fig:mag_chems} and synthesizability metrics discussed in Figure \ref{fig:stability}, we identify the Mn${}_2$Pt${}_{1-z}X_z$Ga system for $X=$ Au, Ir, Ni as a model system for tunably accessing all regions of the magnetic phase diagram.
The Mn${}_2$PtGa endpoint lies in the ASk region of the long-range magnetic phase diagram in Figure \ref{fig:mag_chems}b.
The $X=$Au, Ir, Ni endpoints of the alloy are separated from this region by the ASk/Hx, ASk/Cx, Hx/FiM and Cx/FiM phase boundaries.
Thus by varying the dopant element and composition we expect the alloy to move across the ASk, Hx, Cx, and FiM regions of the magnetic phase diagram, with several critical compositions corresponding to magnetic phase boundaries.
Furthermore, the majority phase Mn${}_2$PtGa is one of the few compositions that thermodynamically favors the tetragonal inverse Heusler structure, which imparts chemical stability to this alloy.
Thus, while Mn${}_2$AuGa for example is metastable in the inverse Heusler structure, a modest amount of Au doping into Mn${}_2$PtGa retains the desired structure.

Figure \ref{fig:pt_doping} shows a quantitative evaluation of the magnetic and chemical behavior of the Mn${}_2$Pt${}_{1-z}X_z$Ga family of alloys for $X = $ Au, Ir, Ni.
Assuming that a solid-solution forms across these compositions, the micromagnetic parameters $K$, $D$ and $A$ must vary smoothly between the alloy endpoints.
For the magnetocrystalline anisotropy $K$, we compare three models for how this quantity varies with alloy composition $z$, as shown in Figure \ref{fig:pt_doping}a.
The simplest model is a linear interpolation $K(z) = K(0) + z(K(1) - K(0))$ which neglects any new  magnetochemical interactions that may appear at intermediate compositions of the alloy.
A more refined model is a cluster expansion (DFT-CE) model, $K = \sum_\alpha J^{(K)}_{\alpha} \prod \sigma_{\alpha}$, where $\alpha$ represent two-, three-, and four-body clusters of sites with chemical occupancy denoted by $\sigma$, and $J^{(K)}_{\alpha}$ are interaction coefficients which capture the contribution of each cluster to the total magnetocrystalline anisotropy.
This model is analogous to a conventional cluster expansion of the total energy\cite{Sanchez1984, VdV2018} and, parametrized using DFT data, captures the impact of distinct chemical environments on the magnetocrystalline anisotropy.
The final model is an explicit DFT calculation of the magnetic anisotropy at select compositions of the alloy using a special quasi-random structure (SQS)\cite{Zunger1990}.
As can be seen in Figure \ref{fig:pt_doping}a, while the DFT-CE and DFT-SQS models consistently indicate a degree of non-linearity in $K(z)$, the deviation from the simple linear interpolation is small.
Thus, in the case of $D$ and $A$, we assume a simple linear interpolation with composition $z$, $D(z) = D(0) + z(D(1) - D(0))$ as accounting for any non-linear contribution to these terms is very computationally expensive and unlikely to substantially affect our conclusions.

In Figure \ref{fig:pt_doping}b, we combine the DFT-CE model for $K(z)$ and linear interpolation models of $A(z)$ and $D(z)$ to evaluate the magnetic phase diagram of Mn${}_2$Pt${}_{1-z}X_z$Ga alloys.
Starting from the ASk region for $z=0$, the wavelength $\lambda = 2\pi A / D$ of the helimagnetic phases increases until the alloy crosses into new regions of magnetic phase space at $z \approx 0.1-0.2$, depending on the choice of $X$ element.
In the case of $X=$ Ir, we expect a transition to Hx-type behavior at $z=0.22$ and easy-axis FiM at $z=0.28$.
For $X=$ Au, the ASk region instead transitions to Cx-type behavior at $z=0.09$ and easy-plane FiM at $z=0.18$.
The $X=$ Ni space contains 3 transitions, from ASk to Cx at $z=0.14$, Cx to easy-plane FiM at $z=0.22$ and easy-plane to easy-axis FiM at $z=0.85$.
Close to the critical $z$-values for these phase transitions, the magnetic behavior is likely to be highly susceptible to mechanical perturbations that would alter $K$ and $D$, such as uniaxial strain along the crystallographic $c$-axis.
Such a strain could move the material to either side of the magnetic phase boundary and thus actuate the magnetic phase transition.

Finally, in Figure \ref{fig:pt_doping}cd we estimate the synthetic accessibility of the  Mn${}_2$Pt${}_{1-z}X_z$Ga alloys at the compositions of interest.
We compute the pseudo-binary binodal and spinodal curves of these alloys to determine the regions of thermodynamic stability and metastability of the solid solution.
We obtain the mixing enthalpy of the $X=$ Au, Ir, Ni alloys from the DFT-CE and DFT-SQS models analogously to the evaluation of the magnetocrystalline anisotropy $K$.
As shown in Figure \ref{fig:pt_doping}c, the $X=$ Ir case shows a small negative mixing enthalpy, indicating that Mn${}_2$PtGa and Mn${}_2$IrGa are likely to be miscible in the tetragonal inverse Heusler structure at all temperatures and compositions.
The $X=$ Au, Ni alloys have a positive mixing enthalpy indicating that these compositions form miscibility gaps.
Figure \ref{fig:pt_doping}d shows the binodal and spinodal regions of these miscibility gaps, assuming ideal solution entropy for the $X$ sublattice.
The compositions of interest $z \approx 0.1-0.2$ are accessible in both $X = $Au, Ni spaces but require a relatively high processing temperature of 800-900 ${}^o$C for initial mixing.
These alloys can then be annealed to induce ordering at lower temperatures as they resist spinodal decomposition above $\approx$ 600 ${}^o$C.
Thus, the Mn${}_2$(Pt,Ir)Ga system is readily miscible and synthetically limited primarily by the large mismatch in the melting temperatures of Ga-rich and Ir-rich precursors.
In contrast, synthesizing the Mn${}_2$(Pt,Au)Ga and Mn${}_2$(Pt,Ni)Ga alloys is likely to require a careful optimization of the processing temperature to form the ordered inverse Heusler structure while suppressing phase separation into the ternary endpoints.

\section*{Discussion}

We have surveyed the magnetic phase space of tetragonal inverse Heusler alloys, focusing on controlling the stability of long-range spin textures such as antiskyrmions.
By constructing solid-solutions between endpoints with the same short-range spin structure, we are able to tune the effective Dzyaloshinskii-Moriya interaction and magnetocrystalline anisotropy in the alloy to vary the long-range magnetic structure.
Specific compositions of this solid solution which place the magnetic interactions near a magnetic phase boundary maximize the magnetoelastic coupling of the material, as here  mechanically-induced perturbations can actuate a magnetic phase transition.

We demonstrate the power of this design principle by identifying the Mn${}_2$Pt${}_{1-z}X_z$Ga alloy with $X = $ Au, Ir, Ni as a candidate for realizing chemically- and mechanically- tunable antiskyrmions.
In this material, we predict that moderate levels of doping ($z \approx 0.1-0.2$) can induce numerous magnetic phase transitions, and couple antiskyrmion stability to small elastic strains at several critical values of $z$.
The specific doping levels where these phase transitions occur are sensitive to the precise evolution of the coarse-grained interaction parameters with composition and short-range order, which we estimate with several state-of-the-art computational methods.
However, independent of these parametrizations, as long as the alloy forms a true solid solution and connects compounds lying on opposite sides of a magnetic phase boundary, a critical value of $z$ is guaranteed to exist.
This fact suggests that tunable magnetic alloys can be designed even without detailed knowledge of their magnetic interaction parameters.
As long as candidate alloy endpoints can be assigned to distinct regions of the magnetic phase diagram shown in Figure \ref{fig:mag_chems}, a critical composition for realizing the magnetic phase transition and large magnetoelastic coupling is guaranteed to exist.

The primary difficulty with implementing this design principle for tunable magnetism is ensuring that the alloy maintains the desired crystal structure and chemical order at intermediate compositions.
We have assumed that the Mn${}_2 XY$ tetragonal inverse Heuslers maintain the structure shown in Figure \ref{fig:structure}a, with negligible mixing between the four sublattices.
While small amounts of intermixing between the sublattices will slightly alter the effective magnetic interactions and would not affect our broad conclusions\cite{Schneeweiss2017}, many Mn${}_2 XY$ compositions are susceptible to substantial disorder and require optimized processing to induce ordering.
For example, mixing between the Mn and $X$ sublattices creates an inversion center in the material and eliminates the Dzyaloshinskii-Moriya interaction that drives the formation of antiskyrmions in this system.
Chemical disorder may also suppress the martensitic transition into the tetragonal phase that is necessary for $D$ and $K$ to be non-zero.
We have identified Mn${}_2 X$Ga for $X = $Ir, Pt, Rh, Ru, Ni and Mn${}_2X$Sn for $X = $Ru, Rh as compositions that are most likely to form the correct structure and chemical order after annealing at moderate temperature as they have a large driving force to order and minimal driving force to decompose.
Conversely, we have found that most Mn${}_2 X$Sn and Mn${}_2 X$In compounds have a large driving force for decomposition and thus are more likely to phase separate if annealed.
While the experimental literature supports our analysis in the case of Mn${}_2$NiGa\cite{Liu2006},  Mn${}_2$PtGa\cite{Nayak2013, Nayak2013b}, Mn${}_2$RhSn\cite{Meshcheriakova2014}, Mn${}_2$PdSn\cite{Xu2016} and Mn${}_2$PtSn\cite{Huh2015, Liu2018}, the apparent order observed in Mn${}_2$PtIn and Mn${}_2$IrSn\cite{Meshcheriakova2014}, and disorder reported in Mn${}_2$RhGa,  Mn${}_2$RuGa and Mn${}_2$RuSn\cite{Kreiner2014} suggest that other processes may need to be considered.
Ultimately, a substantially more detailed understanding of the synthesis process is necessary to quantitatively evaluate the synthesizability of these structures and the feasibility of controlling their chemical order\cite{Kitchaev2016, Bianchini2020}.

\section*{Conclusion}
We have reported a systematic first-principles derivation of tunable magnetic order in the family of Mn${}_2 X Y$ tetragonal inverse Heusler alloys, focusing on designing a robust coupling between room-temperature antiskyrmion stability and  elastic deformation.
To do so, we first constructed a universal phase diagram for the lattice shared by all tetragonal inverse Heuslers, focusing on the long-range modulation of the common ferrimagnetic spin structure.
We characterized the magnetic behavior of all known stable compounds in this space and identified combinations which, when alloyed, may produce magnetic phase transitions as a function of chemical composition and mechanical deformation.
Finally, we performed an in-depth characterization of the magnetic and chemical behavior of  Mn${}_2$Pt${}_{1-z} X_z$Ga with $X = $ Au, Ir, Ni to demonstrate that for $z \approx 0.1-0.2$, this family of alloys can transition between all possible long-range equilibrium spin textures, including antiskyrmions, helices and conical phases.
At several critical compositions, these magnetic phase transitions may be driven by elastic strain, suggesting that this alloy may exhibit giant magnetoelastic coupling and serve as a mechanical actuator for the formation of complex magnetic order.

\section*{Methods}
Electronic structure calculations were performed with the Vienna Ab-Initio Simulation Package (VASP) \cite{Kresse1996a} using the Projector-Augmented Wave method\cite{Kresse1999}.
All magnetic interactions (Figure \ref{fig:mag_chems}, Figure \ref{fig:pt_doping}) were determined using the Perdew-Burke-Ernzerhof (PBE) exchange--correlation functional\cite{PBE1996}, accounting for spin-orbit coupling.
Following previously reported benchmarks, a dense reciprocal-space mesh of 400 $k$-points per \AA${}^{-3}$ was used\cite{Wollmann2015, Sanvito2017, Faleev2017}, making sure that all magnetic calculations of the same chemistry and supercell used exactly the same $k$-point mesh\cite{Kitchaev2020, Schueller2020} and converging the total energy to $10^{-7}$ eV.

The relative stabilities of the ordered and disordered phases (Figure \ref{fig:stability}) were determined using the same computational parameters, but neglecting spin-orbit coupling.
Disordered phases were modeled using special quasi-random structure (SQS)\cite{Zunger1990} representations of the common L$2_{1b}$ (Mn(1)/$X$) and BiF${}_3$ (Mn(1)/Mn(2)/$X$) disorder types in these systems\cite{Kreiner2014}, where each SQS representation is relaxed assuming a ferrimagnetic spin configuration.
To compute global chemical stability within the Mn-$X$-$Y$ chemical spaces (Figure \ref{fig:stability}), we rely on structures reported in the ICSD\cite{ICSD}, Materials Project\cite{Jain2013}, and OQMD\cite{Saal2013} databases, with energies computed using the SCAN exchange-correlation functional\cite{SCAN} as we found that this functional uniquely reproduces the low-$T$ behavior of the known binary phase diagrams and avoids previously reported pathological behavior in e.g. the Pt-based binaries\cite{Decolvenaere2015}.
These chemical stability calculations are converged to $10^{-5}$ eV in total energy and 0.02 eV/\AA \ in forces, and are optimized over likely collinear ferromagnetic and antiferromagnetic configurations for all phases.
The equilibrium phases used to determine formation energies are given Supplementery Data 5.

Magnetic Hamiltonians were obtained following previously described methods for generating a complete basis for quasi-classical spin interactions within a cluster expansion formalism\cite{Drautz2004, Kitchaev2018, Kitchaev2020, VdV2018}.
Briefly, for each symmetrically-distinct group of magnetic sites, we construct interaction basis functions consisting of symmetrized products of spherical Harmonics, e.g. for a pair of spins  $|l_1,l_2;L,M\rangle = 4\pi \sum_{m_1,m_2} c^{l_1,l_2,L}_{m_1,m_2,M}Y^{l_1}_{m_1}(\phi_1, \theta_1)Y^{l_2}_{m_2}(\phi_2, \theta_2)$ where $c^{l_1,l_2,L}_{m_1,m_2,M}$ are Clebsch-Gordan coefficients.
The $L=0$ terms correspond to exchange-interactions, $L=1$ correspond to Dzyaloshinskii-Moriya couplings, and $L=2,4,...$ correspond to magnetocrystalline anisotropies\cite{Kitchaev2020}.
Here, we consider $L = 0$ (exchange) two-spin interactions for first, second and third  nearest-neighbor interactions shown in Figure \ref{fig:structure}b.
$L = 1$ (Dzyaloshinskii-Moriya) terms are included for the nearest-neighbor interaction ($J_1$ pair in Figure \ref{fig:structure}a). 
$ L = 2, 4, ...$ terms are included as an average single-site anisotropy summed over the Mn(1) and Mn(2) sublattices.
A full listing of these basis functions is given in Supplementary Data 1.

We parametrize this Hamiltonian to reproduce the energies obtained from DFT.
We group the basis functions by their $L$-value and fit these groups independently to maximally cancel out numerical noise in the DFT calculations: (1) we fit the $L=0$ interactions to symmetrically distinct collinear spin configurations, (2) the $L=1$ interactions to differences in energy between right- and left- handed helical superstructures of the local ferrimagnetic spin structure, and (3) the $L=2,4,...$ interactions to the energy associated with rotating the ground-state ferrimagnetic spin structure with respect to the crystal axes.
Finally, we fit the coarse-grained magnetic parameters $A$ and $D$ in the low-$T$ limit to the energy of spin helix configurations near the equilibrium wavelength implied by the balance of Dzyaloshinskii-Moriya and exchange forces, where the spin helix energies are evaluated using the parametrized atomistic cluster expansion.

Configurational cluster expansions for the total energy and magnetocrystalline anisotropy (Figure \ref{fig:pt_doping}ac) were constructed and parametrized following standard techniques\cite{VdV2018}, including 2-, 3-, and 4- body interactions. 
Special quasi-random structures (SQS)\cite{Zunger1990} based on these cluster expansions were obtained by Monte Carlo optimization targeting the correlations observed in a random alloy at the desired composition within a 3x3x2 supercell of the conventional cell shown in Figure \ref{fig:structure}a.

To determine the finite-$T$ phase diagram (Figure \ref{fig:mag_pd}c) as well as identify the ground states of the magnetic Hamiltonian (Figure \ref{fig:mag_pd}ab), we rely on auxiliary-spin dynamics Hamiltonian Monte Carlo\cite{Drautz2019, Kitchaev2020} with the No U-Turn Sampling technique\cite{Hoffman2014}, as well as simulated annealing and conjugate-gradient optimization.
The Monte Carlo runs sample 1,000 and 10,000 independent configurations for equilibration and production respectively, where the time between independent samples is estimated from the decay rate of the energy autocorrelation function.
Finite-$T$ runs are performed for an equilibrium helical wavelength equal to 24 unit cells, and using a 24x42x3 supercell of the conventional structure, approximately commensurate with a hexagonal antiskyrmion lattice.

\begin{acknowledgments}
We are grateful to Justin Mayer and Eve Mozur for fruitful discussions.
This research was supported by the Materials Research Science and Engineering Center at UCSB (MRSEC NSF DMR 1720256) through IRG-1.
Computational resources were provided by the National Energy Research Scientific Computing Center, a DOE Office of Science User Facility supported by the Office of Science of the U.S. Department of Energy under Contract No. DE-AC02-05CH11231, as well as the Center for Scientific Computing at UC Santa Barbara, which is supported by the National Science Foundation (NSF) Materials Research Science and Engineering Centers program through NSF DMR 1720256 and NSF CNS 1725797.
\end{acknowledgments}


%

\end{document}